\documentclass[conference]{IEEEtran}
\usepackage{cite}
\usepackage{amsmath,amssymb,amsfonts,mathtools}
\usepackage{bbm, bm}
\usepackage{algorithm}
\usepackage{algorithmic}
\usepackage{graphicx}
\usepackage{textcomp}
\usepackage{xcolor}
\usepackage{subcaption}
\usepackage{siunitx}
\usepackage{hyperref}

\newcommand{\grad}{\nabla}

\newcommand{\bI}{\mathbf{I}}

\newcommand{\bzero}{\mathbf{0}}

\newcommand{\bx}{\mathbf{x}}

\newcommand{\bz}{\mathbf{z}}

\newcommand{\bepsilon}{{\boldsymbol{\epsilon}}}

\newtheorem{remark}{Remark}

\def\BibTeX{{\rm B\kern-.05em{\sc i\kern-.025em b}\kern-.08em
    T\kern-.1667em\lower.7ex\hbox{E}\kern-.125emX}}
    
\definecolor{mehdi}{RGB}{0,0,250}

\definecolor{Samad}{RGB}{0,250,0}

\definecolor{Matti}{RGB}{250,0,0}

\graphicspath{ {./Figures/} } 
    
\begin{document}

\title{Denoising Diffusion 
Probabilistic  
Models for Hardware-Impaired  Communications 
}

\author{
    \IEEEauthorblockN{Mehdi Letafati, Samad Ali, and Matti Latva-aho\\}
    \IEEEauthorblockA{\centering
    \begin{tabular}{c}
    \textit{Centre for Wireless Communications,}
    \textit{University of Oulu,}
      Oulu, Finland\\
    \texttt{\{mehdi.letafati, samad.ali, matti.latva-aho\}@oulu.fi}
	\end{tabular}
    \vspace{-1.0\baselineskip}}
\vspace{-1.5\baselineskip}}

\IEEEaftertitletext{\vspace{-2.5\baselineskip}}

\maketitle

\begin{abstract}
Generative  AI 
has received significant attention among a spectrum of diverse   industrial  and  academic domains, thanks to the magnificent results achieved from  deep generative models such as generative pre-trained transformers (GPT) and diffusion models. 
In this paper, we explore the applications of denoising diffusion probabilistic models (DDPMs)  in wireless communication  systems under   practical assumptions such as hardware impairments (HWI), low-SNR regime, and quantization error.   
Diffusion models are a new class of state-of-the-art generative models that 
have  already showcased notable success with  some of the popular examples  
by OpenAI\footnote{https://openai.com/dall-e-2}  and  
Google Brain\footnote{https://imagen.research.google/}.  
The intuition behind DDPM is to decompose the data  generation process over small “denoising” steps.   
Inspired by this, we propose using denoising diffusion model-based   receiver for a practical wireless communication scheme, while providing \emph{network resilience} in  low-SNR regimes, non-Gaussian noise,  different HWI levels, and quantization error. We evaluate the reconstruction  performance of our scheme in terms of 
mean-squared error (MSE) metric.   
Our results  show that more than $25$ dB improvement in MSE  is achieved compared to deep neural network (DNN)-based receivers.  
We also highlight \emph{robust out-of-distribution} performance under non-Gaussian noise. 
\end{abstract}


\vspace{-0.5mm}
\section{Introduction}   
\vspace{-0.5mm}
The emergence of innovative approaches in generative artificial intelligence (GenAI)  has lead to the development of  novel  ideas for  
AI-based systems 
\cite{Petar}. 
At the same time, from data communication and networking perspective, ``connected intelligence'' is envisioned  as the most significant driving force in the sixth generation (6G) of communications---machine learning (ML) and AI algorithms are envisioned to be widely incorporated into 6G networks, realizing  “AI-native” wireless systems \cite{3gpp,  hexaX}.   
This underscores the need for novel AI/ML-based solutions to be tailored for the emerging communication scenarios \cite{twelve_6G, DGM_Mag}.    

Most of the research carried out so far on AI-native wireless has been focused on ``discriminative models’’. One can consider  \cite{DeepRx} and \cite{DNN} as two seminal papers 
that have attracted remarkable attention in both  academia and industry.  From a very high-level perspective, the  goal of such models is to simply  learn the ``boundaries’’ between classes or latent spaces of high-dimensional signals. On the other hand, ``generative models'' are aimed to learn the ``representations'' 
of signals, and generate the desired samples accordingly. In this paper, we  take a radically different approach, and our aim is to unleash the power of GenAI for wireless systems.

One of the recent breakthroughs in GenAI is the evolution of diffusion models, as the new state-of-the-art family of generative models \cite{DM_Ho}. 
It has lead  to     unprecedented   results  in different  applications  such as computer vision, natural language processing (NLP),     
and medical imaging \cite{DM_MRI}.     
The key idea behind  diffusion models is that \emph{if we could develop an ML model that can learn the \emph{systematic decay of information} 
then it should be possible to ``reverse'' the process and recover the information back from the noisy/erroneous data.} The close underlying  relation between the key concepts  on how diffusion models work and the problems in wireless communications  has motivated us to carry out this research.  
Notably, \emph{the incorporation  of diffusion models  into wireless communication problems is still in its infancy, hoping that this paper  would shed light on some of the possible directions.}

Ongoing research  on diffusion models  encompasses both theoretical advancements and practical applications  across  different domains of computer science society. However,   there have been only a few papers in wireless  communications literature that have started looking into the potential merits  of  diffusion  models for wireless systems \cite{DGM_Mag, CDDM, DM_for_E2EComm, CGM_ChanEst, hybrid}.   
The authors in \cite{DGM_Mag} study a workflow for utilizing  diffusion models in wireless network management.  They exploit the flexibility and exploration ability of diffusion models for generating  contracts using diffusion models in mobile AI-generated content services as a use-case.  
Diffusion  models  are utilized in \cite{DM_for_E2EComm} to generate synthetic channel realizations.  
The authors tackle the  problem of  
differentiable channel model within the  
training  process of  end-to-end ML-based communications. 
The results highlight the performance  of diffusion models as an alternative to generative adversarial network (GAN)-based schemes. The authors show that GANs experience unstable training and less diversity in generation performance, while diffusion models  maintain a more stable training process  and a better generalization in inference.  
Noise-conditioned neural networks are employed in \cite{CGM_ChanEst}  for channel estimation in multi-input-multi-output (MIMO) wireless communications. The authors   
employ RefineNet neural architecture  
and run posterior sampling to generate channel estimations based on the pilot signals observed  at the receiver.   
The results in \cite{CGM_ChanEst}  highlight a competitive performance for both in- and out-of-distribution (OOD) scenarios   compared to GANs.    
In \cite{hybrid},  deep learning-based  joint source-channel coding (Deep-JSCC) is combined with diffusion models to  complement  digital communication schemes with a generative component.  The results indicate that the perceptual quality of reconstruction can be improved by employing diffusion models.

Despite the close relation between the ``denoise-and-generate'' characteristics of diffusion models  and the problems in communication theory, to the best of our knowledge, there is only one preprint \cite{CDDM} in the literature that studies the  application of denoising diffusion models  in wireless to help improve the receiver's  performance in terms of noise removal.   
The authors utilize a diffusion model  and call it channel denoising diffusion model (CDDM).  However, the paper has some drawbacks,  which need further considerations.  Specifically, i)  
the authors do not evaluate the  performance of their CDDM module  under realistic scenarios such as hardware impairments (HWI), low-SNR regimes, and non-Gaussian noise. 
Rather, their goal is to simply compensate for the channel noise and equalization errors.    
ii) 
To reconstruct and generate samples,  the authors  implement another neural decoder block at the receiver in addition to CDDM. 
However, having two different ML models, each  maintaining  a   distinct objective function  can impose computational overhead  to the network.  
iii) The proposed scheme in \cite{CDDM} relies on the channel state information (CSI) knowledge at the diffusion module, and the transmitter 
has to feed back the CSI data to the receiver, which can cause communication  overhead and might not be aligned with communication standards.

\subsubsection*{\textbf{Our Work}}
In this paper, we   study the implementation  of denoising diffusion probabilistic models (DDPM),  proposed by  Ho \emph{et al.} in 2020 \cite{DM_Ho}, for a practical  wireless communication  systems with HWIs.      
The key idea of DDPM is to decompose the data  generation process into subsequent  “denoising” steps and then gradually generating the desired samples out of noise.  
Inspired by the recent visions on AI-native wireless for 6G \cite{hexaX}, our general idea in this paper is that instead of designing a communication system which  avoids HWI and estimation/decoding errors,  we can train the network to  
handle such distortions,   
aiming to introduce   \emph{native resilience} for wireless AI \cite{hexaX, 3gpp, twelve_6G}.  

In our proposed approach, a DDPM 
is employed at the receiver side (without relying on any other conventional  autoencoder in contrast to \cite{CDDM})  to enhance  the resilience of the wireless system   
against  practical non-idealities  such as HWI and quantization error.  
Our DDPM  is parameterized  by a neural network (NN) comprised  of conditional linear layers.   
Inspired by the Transformer paper (Vaswani \emph{et al.,} 2017 \cite{attention}), we only employ one   model for  the entire denoising time-steps   
by incorporating the time  embeddings  into the model.  
After the diffusion model is trained to generate data samples out of noise,  
the receiver runs the reverse diffusion process algorithm, starting from the distorted received  signals, to reconstruct the transmitted data.    
We demonstrate the  \emph{resilience} of our  DDPM-based approach in different scenarios, including  low-SNR regimes, non-Gaussian noise,  different HWI levels, and quantization error, which were not addressed in \cite{CDDM}.   
We also evaluate the mean-squared error (MSE) of our scheme, highlighting  more than  $25$ dB performance improvement compared to the  deep neural network (DNN)-based receiver of  \cite{DNN} as one of the promising benchmark  designs in ML-based  communications systems.

In the following sections,  we first introduce the concept of DDPMs  together with the main formulas in Section \ref{sec:DDPM}. Our system model is introduced in Section \ref{sec:SysMod}, where  we provide the generic formulations and the details of the neural architecture and  algorithms.  Then, we study  numerical evaluations in Section \ref{sec:Eval}, and conclude the paper in  Section \ref{sec:concl}.\footnote{\emph{Notations:} 
Vectors and matrices are represented, respectively,  by bold lower-case and upper-case symbols. $|\cdot|$ and $||\cdot ||$ respectively denote the  absolute value of a scalar variable and the $\ell_2$ norm of a vector. Notation $\mathcal{N}(\mathbf{x}; \boldsymbol{\mu}, \mathbf{\Sigma})$  stands for the multivariate normal distribution  with mean vector $\boldsymbol{\mu}$ and covariance matrix $\mathbf{\Sigma}$ for a random vector $\mathbf{x}$. Similarly, complex normal distribution with the corresponding mean vector  and covariance matrix is denoted by $\mathcal{CN}(\boldsymbol{\mu}, \mathbf{\Sigma})$. Moreover, the expected value of a random variable (RV) is denoted by $\mathbb{E}\left[\cdot\right]$   Sets are denoted by calligraphic symbols.  $\bm 0$ and $\bf I$ respectively show all-zero vector and identity matrix. Moreover, $[N]$,  (with $N$ as integer) denotes the set of all integer values from $1$ to $N$, and $\mathsf{Unif}[N]$, $N > 1$, denotes discrete uniform distribution  with samples between $1$ to $N$.   
}

\vspace{-1mm}
\section{Preliminaries on DDPMs}\label{sec:DDPM}
\vspace{-0.5mm}
Diffusion models are a new class of generative models that are inspired by non-equilibrium thermodynamics \cite{DM_Ho}. They  consist of two diffusion processes, i.e., the forward and the  reverse  process. During  the forward diffusion steps, random perturbation  noise is purposefully added to the original data. Then in a reverse process, DDPMs learn to construct the desired data samples out of noise.

 Let ${\bf x}_0$  be a data sample from some  distribution ${ q}({\bf x}_0)$. 
 For a finite number, $T$, of time-steps, the forward diffusion process
 $ q({\bf x}_t|{\bf x}_{t-1})$ 
 is defined by adding  Gaussian noise at each time-step 
$t \in [T]$  according to a known ``variance schedule'' $0 < \beta_1 <  \beta_2 < \cdots <  \beta_T < 1$. 
 This can be formulated as 
\begin{align}
    q(\mathbf{x}_t \vert \mathbf{x}_{t-1}) 
    & \sim \mathcal{N}(\mathbf{x}_t; \sqrt{1 - \beta_t} \mathbf{x}_{t-1}, \beta_t\mathbf{I}),  \label{eq:fwd_diffusion}  \\
q(\mathbf{x}_{1:T} \vert \mathbf{x}_0) 
& = \prod^T_{t=1} q(\mathbf{x}_t \vert \mathbf{x}_{t-1}). \label{eq:diffusion_eqn}
\vspace{-1mm}
\end{align} 
Invoking \eqref{eq:diffusion_eqn}, the data sample  gradually loses its distinguishable features as the time-step goes on, where with $T\!\rightarrow \!\infty$, ${\bf x}_T$ approaches an isotropic Gaussian distribution with covariance matrix ${\bf \Sigma}\!=\!\sigma^2\mathbf{I}$ for some $\sigma\!>\!0$ \cite{DM_Ho}. According to \eqref{eq:fwd_diffusion},  each new sample at time-step  $t$ can be drawn from a conditional Gaussian distribution with  mean vector  ${\mathbf \mu}_t = \sqrt{1 - \beta_t} \mathbf{x}_{t-1}$ and covariance matrix ${\bf \Sigma}^2_t = \beta_t \bf I$. Hence, the forward process is realized  by sampling from a Gaussian noise  $\bm{\epsilon}_{t-1} \sim {\cal N}(\bf 0, I)$  and setting 
\begin{align}\label{eq:fwd_sample_gen_diffusion}
	{\bf x}_t = \sqrt{1-\beta_t}&{\bf x}_{t-1} +\sqrt{\beta_t} {\bm{\epsilon}}_{t-1}.
\end{align} 
A useful property for  the forward  process in \eqref{eq:fwd_sample_gen_diffusion} is that we can sample ${\bf{x}}_t$ 
at any arbitrary time step $t$  in a closed-form expression,  through recursively applying the reparameterization trick from  ML literature \cite{reparam_ML}.   
This results in 
\begin{align} 
\mathbf{x}_t  &= \sqrt{\bar{\alpha}_t}\mathbf{x}_0 + \sqrt{1 - \bar{\alpha}_t}\bm{\epsilon}_0,  \label{eq:xt_vs_x0} \\ 
    q({\bf x}_t|{\bf x}_0)&\sim\mathcal{N}\left({\bf x}_t;\sqrt{\bar{\alpha}_t}{\bf x}_0,(1-\bar{\alpha}_t)\mathbf{I}\right),\label{eq:xt_vs_x0_dist}
\end{align}
where $\bar{\alpha}_t\!=\!\prod_{i=1}^t(1-\alpha_i)$ and $\alpha_t=1-\beta_t$  \cite{reparam_ML}. 

Now the problem is to  reverse the process in \eqref{eq:xt_vs_x0} and sample from 
$q(\mathbf{x}_{t-1} \vert \mathbf{x}_t)$, so that we regenerate  the true samples from
$\mathbf{x}_T$.  
According to \cite{DM_Ho}, for $\beta_t$ 
small enough,  $q(\mathbf{x}_{t-1} \vert \mathbf{x}_t)$  also follows Gaussian distribution $\forall t \in [T]$. However,  we cannot  estimate the distribution, since  it requires knowing the distribution of all possible data samples (or equivalently exploiting  the entire dataset). 
Hence,  to approximate the conditional probabilities and  run the reverse diffusion process, we need to learn a probabilistic  model $p_{\bm \theta}(\mathbf{x}_{t-1} \vert \mathbf{x}_t)$ that  is parameterized by ${\bm \theta}$.   
According to the above explanations, the following expressions can be written  
\vspace{-2mm}
\begin{align} 
p_{\bm \theta}(\mathbf{x}_{t-1} \vert \mathbf{x}_t) &\sim \mathcal{N}(\mathbf{x}_{t-1}; \boldsymbol{\mu}_{\bm \theta}(\mathbf{x}_t, t), \mathbf{\Sigma}_{\bm \theta}(\mathbf{x}_t, t)),  \label{eq:rev_proc_dist_conditional} 
\\ 
 p_{\bm \theta}(\mathbf{x}_{0:T}) &= p(\mathbf{x}_T) \prod^T_{t=1} p_{\bm \theta}(\mathbf{x}_{t-1} \vert \mathbf{x}_t).  \label{eq:rev_proc_dist_all} 
\end{align} 
Hence,  the problem simplifies  to learning the mean vector  $\boldsymbol{\mu}_{\bm \theta}(x_t,t)$ and the covariance matrix  $\mathbf{\Sigma}_{\bm \theta}(x_t,t)$ 
for the  probabilistic model $p_{\bm \theta}(\cdot)$, where an NN can be trained to approximate (learn) the reverse process.

Before proceeding with the details of learning the reverse diffusion process, we note that if we condition the reverse process  on ${\bf x}_0$, this conditional probability becomes  tractable.    
The intuition behind this could be explained as follows.  \emph{A painter (our generative model) requires a reference image ${\bf x}_0$ to be able to gradually  draw a picture.}  Hence, when we have  ${\bf x}_0$ as a reference, we can take a small step backwards from noise to generate the data samples. Then, the reverse step is formulated as $q({\bf x}_{t-1}|{\bf x}_t,{\bf x}_0)$.  
Mathematically speaking,  
we can  derive $q({\bf x}_{t-1}|{\bf x}_t,{\bf x}_0)$ using   Bayes rule as follows. 
\begin{align}
    q(\mathbf{x}_{t-1} \vert \mathbf{x}_t, \mathbf{x}_0) &\sim \mathcal{N}(\mathbf{x}_{t-1}; \hspace{1.5mm} {\tilde{\boldsymbol{\mu}}}(\mathbf{x}_t, \mathbf{x}_0, t), {\tilde{\beta}_t} \mathbf{I}),   \label{eq:rev_conditioned_on_x0}
\end{align}
where 
\begin{align}
    {\tilde{\boldsymbol{\mu}}}(\mathbf{x}_t, \mathbf{x}_0, t)
    &=\frac{\sqrt{\alpha_t}(1-\bar{\alpha}_{t-1})}{1-\bar{\alpha}_t}{\bf x}_t + \frac{\sqrt{\bar{\alpha}_{t-1}}
    \beta_t 
    }{1-\bar{\alpha}_t}{\bf x}_0, 
    \text{\hspace{2mm}and}
    \label{eq:mu_tilde}  \\ 
   {\tilde{\beta}_t} &=\frac{
   1-\bar{\alpha}_{t-1}}{1-\bar{\alpha}_t} \beta_t.\label{eq:beta_tilde}
\end{align}
Invoking \eqref{eq:beta_tilde}, one can infer that the covariance matrix in \eqref{eq:rev_conditioned_on_x0} has no learnable parameter.  
Hence, we simply need to learn the  mean vector ${\tilde{\boldsymbol{\mu}}}(\mathbf{x}_t, \mathbf{x}_0, t)$. 
To further simplify \eqref{eq:mu_tilde}, we note that  thanks to the reparameterization trick and with a similar approach to  \eqref{eq:xt_vs_x0},  we can express  ${\bf x}_0$ as follows.    
\begin{align}
    \mathbf{x}_0 = \frac{1}{\sqrt{\bar{\alpha}_t}}(\mathbf{x}_t - \sqrt{1 - \bar{\alpha}_t}\bm{\epsilon}_t). \label{eq:x0_vs_xt}
\end{align}
Substituting ${\bf x}_0$ in \eqref{eq:mu_tilde} by \eqref{eq:x0_vs_xt} results in 
\begin{align}
    \begin{aligned}
\tilde{\boldsymbol{\mu}}(\mathbf{x}_t, \mathbf{x}_0, t) = 
{\frac{1}{\sqrt{\alpha_t}} \Big( \mathbf{x}_t - \frac{1 - \alpha_t}{\sqrt{1 - \bar{\alpha}_t}} \bm{\epsilon}_t \Big)}.
\end{aligned}
\end{align}
Now we can learn the conditioned probability distribution $p_{\bm \theta}(\mathbf{x}_{t-1} \vert \mathbf{x}_t)$
of the reverse diffusion process 
by training a NN  that approximates $\tilde{\boldsymbol{\mu}}(\mathbf{x}_t, \mathbf{x}_0, t)$.    
Therefore,  we simply need to set the approximated mean vector  $\boldsymbol{\mu}_{\bm \theta}(\mathbf{x}_t, t)$ to have the same form as the target mean vector $\tilde{\boldsymbol{\mu}}(\mathbf{x}_t, \mathbf{x}_0, t)$.  
Since $\mathbf{x}_t$  is known at time-step $t$, we can reparameterize the NN to make it approximate $\bm{\epsilon}_t$ 
 from the input $\mathbf{x}_t$. Compiling these facts  results in the following expression for $\boldsymbol{\mu}_{\bm \theta}(\mathbf{x}_t, t)$  
 \begin{align}\label{eq:mu_theta_to_learn}
\boldsymbol{\mu}_{\bm \theta}(\mathbf{x}_t, t) &=  {\frac{1}{\sqrt{\alpha_t}} \Big( \mathbf{x}_t - \frac{1 - \alpha_t}{\sqrt{1 - \bar{\alpha}_t}} \bm{\epsilon}_{\bm \theta}(\mathbf{x}_t, t) \Big)}, 
\end{align}
where $\bm{\epsilon}_{\bm \theta}(\mathbf{x}_t, t)$ denotes our NN.  

We now define the loss function $\mathcal{L}_t$ 
aiming  to minimize the difference between $\boldsymbol{\mu}_{\bm \theta}(\mathbf{x}_t, t)$ and  $\tilde{\boldsymbol{\mu}}(\mathbf{x}_t, \mathbf{x}_0, t)$.   
\begin{align} 
\hspace{-2.5mm}
\mathcal{L}_t &=  
{\mathbb{E}}_{\begin{subarray}{l}t\sim {\mathsf{Unif}}[T]\\ {\bf x}_0\sim q({\bf x}_0) \\ \bm{\epsilon}_0\sim \mathcal{N}(0,\bf{I})\\ \end{subarray}}
\Big[\|\bm{\epsilon}_t - \bm{\epsilon}_{\bm \theta}(\mathbf{x}_t, t)\|^2 \Big]  \nonumber \\  
&=  {\mathbb{E}}_{\begin{subarray}{l}t\sim {\mathsf{Unif}}[T]\\ {\bf x}_0\sim q({\bf x}_0) \\ \bm{\epsilon}_0\sim \mathcal{N}(0,\bf{I})\\ \end{subarray}}  \Big[\|\bm{\epsilon}_t - \bm{\epsilon}_{\bm \theta}(\sqrt{\bar{\alpha}_t}\mathbf{x}_0 + \sqrt{1 - \bar{\alpha}_t}\bm{\epsilon}_t, t)\|^2 \Big]. \label{eq:loss_func}
\end{align} 
Invoking \eqref{eq:loss_func}, 
at each time-step $t$,  the  DDPM model   
takes $\mathbf{x}_t$ as input  and returns the distortion components $\bm{\epsilon}_{\bm \theta}(\mathbf{x}_t,t)$. Moreover, $\bm{\epsilon}_t$ denotes the  diffused noise term  at time step $t$.

\section{System Model and Proposed Scheme}\label{sec:SysMod}   

\subsection{Problem Formulation}\label{subsec:ProblemFormula}  
Consider a point-to-point communication system 
with non-ideal transmitter and receiver  hardware.   We denote by $s_k \in  \mathbb{C}$, the $k$-th element, $k \in [K]$,  in the batch (with size $K$) of unit-power  
data samples  that are supposed to be transmitted over the air.
Wireless channel between the communication entities follows block-fading model, and is represented by a  complex-valued scalar $h_k \in \mathbb{C}, \forall k \in [K]$, taking   independent realizations within  each  coherence block. 
The corresponding  received signal $y_k$  
under non-linear HWIs  is  formulated by 
\begin{align}\label{eq:HWI_scalar}
{y_k}= {h}_k (\sqrt{p} {s}_k + {\eta}^{t}_k) + \eta_{k}^{r} + n_k, 
\end{align}
where $p$ denotes the transmit power,  $\eta_{k}^t \sim \mathcal{CN}({0},\kappa^t p)$ is the distortion noise caused by the transmitter hardware with the corresponding impairment level $\kappa^t$. 
Moreover,  
$\eta_{k}^{r}$ reflects the hardware distortion at the receiver with $\kappa^r$ showing the level of impairment at the 
receiver hardware. 
Notably, ${\eta^r_k}$ is conditionally Gaussian, given the channel realization  ${h}_k$, and  $\eta^r_k\sim \mathcal{CN}(0,\kappa^r p|h_{k}|^2)$.\footnote{This is an experimentally-validated model for HIs, which is widely-adopted in wireless communication literature  \cite{HI}.} 
To further explain \eqref{eq:HWI_scalar}, we emphasize that  according to \cite{HI},  the distortion noise caused at each radio frequency (RF) device  is proportional to its signal power. In addition to the receiver noise $n_k$ that models random fluctuations in the electronic circuits of  the receiver,  a fixed portion of the information signal is turned into \emph{distortion noise} due to 
inter-carrier interference induced by phase noise, leakage from the mirror subcarrier under I/Q imbalance, nonlinearities in power amplifiers, etc. \cite{HI}.   

\begin{remark}
The power of distortion noise is proportional to the signal power $p$ and the channel gain $|{h_k}|^2$.  
According to \cite{HI}, HWI levels, $\kappa^t$ and $\kappa^r$,  characterize the proportionality coefficients and are related to the error vector magnitude (EVM). 
Following \cite{HI}, we consider $\kappa$-parameters in the range $[0,0.15^2]$ in our simulations, where smaller values imply less-impaired  transceiver hardware.
\end{remark}

\begin{remark}
The additive noise term $n_k \sim \mathcal{CN}(0, \delta^2)$ in \eqref{eq:HWI_scalar} could be interpreted as the aggregation of   independent receiver noise   and interference   from simultaneous transmissions of other nodes. In this paper,  for the sake of notation  brevity,  we have used a common notation $n_k$ for the aggregate effect of  interference and noise.  Investigation of interference signals will be addressed in  our subsequent  works.  
\end{remark} 

Adhering to the
``batch-processing'' nature of AI/ML algorithms and AI/ML-based wireless communications  simulators \cite{Sionna}, 
we formulate data signaling expressions in matrix-based format.     
We define a batch of data samples by ${\bf{s}} \overset{\Delta}{=}  \left[s_1, \cdots, s_K\right]^{\sf T}$ with the underlying distribution $\mathbf{s} \sim q(\mathbf{s})$.  We also define  the corresponding channel realizations vector as ${\bf h} \overset{\Delta}{=} \left[ h_{1}, \ldots, h_K \right]^{\sf T}$. Similarly, ${\bm{\eta}}^r \overset{\Delta}{=} \left[  \eta^r_1,   \ldots,  \eta^r_K \right]^{\sf T} \in \mathbb{C}^{K}$, where  
$\bm{\eta}^r$ is conditionally Gaussian given the channel realizations  $\{{h}_k\}_{k \in [K]}$, i.e., $\bm{\eta}^r \sim \mathcal{CN}({\bf 0}_K, \kappa^r p {\mathbf{G}})$, with $\mathbf{G} = \mathsf{diag}(\mathbf{g})$, where $\mathbf{g} \overset{\Delta}{=} \left[ |h_1|^2, \ldots, |h_K|^2 \right]^{\sf T}$. 
We also define  ${\bm \eta}^t \overset{\Delta}{=} \left[  \eta^t_1,   \ldots,  \eta^t_K \right]^{\sf T} \in \mathbb{C}^{K}$ with   ${\bm \eta}^t \sim \mathcal{CN}({{\bf{0}}_K},\kappa^t p {\bf{I}}_K)$, and   
${\mathbf n} \sim \mathcal{CN}({\bf 0}_K, \sigma^2{\bf I}_K)$.  
Then we can rewrite the batch $\mathbf{y} \in \mathbb{C}^K$ of received samples, which is given below. 
\begin{align}\label{eq:received_y_tensor}
    \mathbf{y} =  \sqrt{p} \hspace{1mm} \mathbf{H}  \mathbf{s} + \bm{\zeta}, 
    \vspace{-2mm}
\end{align}
where $\mathbf{H} =  \hspace{1mm} \mathsf{diag}(\bf h)$, and ${\bm \zeta} \overset{\Delta}{=}  \mathbf{H}\bm{\eta}^t + \bm{\eta}^r + \mathbf{n}$ represents  the  
effective Gaussian noise-plus-distortion  given the channel vector ${\bf h}$ with the conditional distribution $\bm{\zeta} \vert{\bf h} \sim \mathcal{CN}({{\bf 0}}_K,{\bf \Sigma})$ and  
the covariance matrix ${\bf \Sigma}$ given by
${\bf \Sigma} = p (\kappa^t + \kappa^r) \mathbf{G} + \sigma^2 \mathbf{I}_K$. 
Since NNs can only process real-valued inputs,  
we  map complex-valued symbols to real-valued tensors, and rewrite the received signals in \eqref{eq:received_y_tensor}
by stacking the real and imaginary components. This results in the following expression.   
\begin{align} \label{eq:received_z_tensor} 
\mathbf{y}_{\sf r} = \tilde{\mathbf{H}} \mathbf{x} + \bm{\nu}, 
\end{align}
where  ${\bf y}_{\sf r}\hspace{-1mm}=\hspace{-1mm}\begin{bmatrix} \Re\{{\bf y}\} \\ \Im\{{\bf y}\} \end{bmatrix}$, ${{\bf x}}\hspace{-1mm}=\hspace{-1mm} \sqrt{p} \begin{bmatrix}  \Re\{{\bf s}\}  \\ \Im\{{\bf s}\} \end{bmatrix}$, ${\tilde{{\bf H}}}\hspace{-1mm}=\hspace{-1mm}\begin{bmatrix} \Re\{{\bf H}\} & \hspace{-1mm} -\Im\{{\bf H}\} \\ \Im\{{\bf H}\} & \hspace{-1mm} \Re\{{\bf H}\}\end{bmatrix}$, and ${\bm \nu}\hspace{-1mm}=\hspace{-1mm}\begin{bmatrix} \Re\{\bm{\zeta}\} \\ \Im\{\bm{\zeta}\} \end{bmatrix}$. 
For the completeness of our  formulations,  we  note that  the effective  noise-plus-distortion term $\bm{\zeta}$ in \eqref{eq:received_z_tensor} is conditionally circulary symmetric given the channel realizations vector. Hence,  the covariance matrix of ${\bm \nu}$  could be written as  
${\bf C}=\frac{1}{2}\begin{bmatrix} \Re\{{\bf \Sigma}\} & -\Im\{{\bf \Sigma}\} \\ \Im\{{\bf \Sigma}\} & \Re\{{\bf \Sigma}\} \end{bmatrix}\in \mathbb{R}^{2K \times 2K}$.

\subsection{DDPM Solution:  Algorithms \& Neural Architecture}\label{subsec:Algorithms} 
In this subsection, we exploit the DDPM framework  
for our 
hardware-impaired  wireless  system.   
First, a NN is trained  with the aim  of learning  hardware and channel distortions.  Then, exploiting the so-called ``denoising'' capability of our DDPM  helps us reconstruct samples from the original data distribution, by removing imperfections and distortions from the batch of distorted  received signals,  $\mathbf{y}_{\sf r}$.  
Hence, the combined effect of HWI and communication channel is taken into account. 
In brief, the idea here is to first make our system learn how to gradually remove the purposefully-injected  noise from batches of  noisy samples ${\mathbf{x}}_T$.  
Then  in the inference (sampling) 
phase, we start  from the batch of received signals 
in \eqref{eq:received_z_tensor}, run the DDPM framework to remove  the hardware and channel distortions   
and reconstruct  original samples $\mathbf{x}$.   

\subsubsection{Neural Network Architecture}
 Generally speaking, the implemented NN is supposed  to take as input the tensor of noisy signals at a particular time step, and  approximate   distortions-plus-noise as the output.  This approximated distortion would be a tensor with the same size as the input tensor.  Hence, the NN should output  tensors of the same shape as its input.   
Accordingly,   our DDPM  model  is parameterized  by a  NN, $\bm{\epsilon}_{{\bm \theta}}(\cdot, t)$, with $3$ conditional hidden layers (with softplus activation functions), each of which has $128$
neurons conditioned on $t$.    
The output layer is a simple  linear layer with the same size as the input.   
To enable  employing only  one neural  model for  the entire denoising time-steps,  the hidden layers are conditioned on $t$  by multiplying the embeddings of time-step.   
More specifically, inspired by the Transformer architecture  (Vaswani et al., 2017 \cite{attention}), we share the   parameters of the NN across time-steps via  incorporating the embeddings of time-step  into the  model.     
Intuitively, this makes the neural network ``know''  at which particular time-step it is operating for every sample in the batch.  
We  further emphasize that the notion of time-step equivalently corresponds to the level of residual  noise-plus-distortion in the batch of received signals. This  will be verified in our simulation results in Section \ref{sec:Eval} (Fig. \ref{fig:train}), where   data samples with different levels of ``noisiness'' can be observed  at  different time-steps $t \in [T]$.

\subsubsection{Training and Sampling Algorithms}  
Our diffusion model is trained based on the loss function given in \eqref{eq:loss_func},  using the MSE between the true and the predicted distortion noise. 
In  \eqref{eq:loss_func}, $\mathbf{x}_0 \sim q(\mathbf{x}_0)$  stands for  data samples from a training set  with unknown and possibly  complex 
underlying 
distribution $q(\cdot)$,   
 and $\bm{\epsilon}_{\bm \theta}(\mathbf{x}_t, t)$ denotes the approximated noise-plus-distortion  at the output of the NN. 
We note  that  $\bar{\alpha}_t$ in \eqref{eq:loss_func}  is a function of the  variance scheduling $\beta_t$ that we design. Therefore,  $\bar{\alpha}_t$'s are known and can be calculated/designed beforehand.  
This leads to the fact that by properly designing the variance scheduling for our forward diffusion process, we can optimize a wide range of desired 
loss functions  
$\mathcal{L}_t$ during training.     
Hence, intuitively speaking,  the NN would be able to ``see''  different structures of the distortion noise  during its  training,  making it robust against a wide range of distortion levels for  sampling. 
The training process of the proposed DDPM  is  summarized in Algorithm \ref{alg:trainAlg}. 
\begin{algorithm}[t]
\small
\hspace*{0.02in} {\bf {Hyper-parameters:}}
	{Number of time-steps $T$, neural architecture $\boldsymbol{\epsilon}_{\bm \theta}(\cdot, t)$, variance schedule   $\beta_t$, and $\bar{\alpha}_t, \forall t \in [T]$.} \\
    \hspace*{0.02in} {\bf {Input:}}
	{Training samples $\mathbf{x}_0 \sim q(\mathbf{x}_0)$  from a dataset $\cal S$.} \\
	\hspace*{0.02in} {\bf {Output:}} {Trained neural model for DDPM.}
	\caption{\small Training algorithm of DDPM}
	\label{alg:trainAlg}
	\begin{algorithmic}[1] 
 \small
    \WHILE {the stopping criteria are not met}
    \STATE Randomly sample $\mathbf{x}_0$ from $\cal S$
    \STATE Randomly sample $t$ from $\mathsf{Unif}[T]$ 
    \STATE Randomly sample $\boldsymbol{\epsilon}$ from $\mathcal{N}(\mathbf{0}_{2K},\mathbf{I}_{2K})$ 
      \STATE Take gradient descent step on
      \STATE $\qquad {\grad}_{\boldsymbol{\theta}} \left\| \bepsilon - \bepsilon_{\bm \theta}(\sqrt{\bar\alpha_t} \bx_0 + \sqrt{1-\bar\alpha_t}\bepsilon, t) \right\|^2$
    \ENDWHILE
	\end{algorithmic}
\end{algorithm}
We take a random sample $\mathbf{x}_0$
 from the training set  $\cal S$; 
A  random time-step $t \sim \mathsf{Unif}[T]$ is  embedded into the conditional  model;  
We also sample a noise vector $\boldsymbol{\epsilon}$ (with the same shape as the input) from  normal  distribution, and distort  the input by this noise  according to the desired   
``variance scheduling''   $\beta_t$; The NN is trained to estimate the noise vector in the distorted data  
$\mathbf{x}_t$.

The sampling process  of our scheme   
is summarized in Algorithm \ref{alg:sampling}. 
We start from the received batch of distorted signals  $\mathbf{y}_{\sf r}$, and then  iteratively denoise it using  the trained  NN. 
More specifically, starting from $\mathbf{y}_{\sf r}$, for each time step $t\in\{T, T-1, \ldots, 1\}$,  the NN outputs $\bm{\epsilon_{\bm \theta}}(\mathbf{x}_t, t)$ to approximate  the residual noise-plus-distortion within the batch of received signals.     
A sampling algorithm is then run as expressed in step $4$ of algorithm \ref{alg:sampling}, in order  to sample $\mathbf{x}_{t-1}$. The process is executed for $T$ time-steps until $\mathbf{x}_0$ is reconstructed ultimately.

\begin{figure}
\vspace{-5mm}
\begin{algorithm}[H]
\small
  \caption{\small Sampling  algorithm of DDPM} \label{alg:sampling}
  \begin{algorithmic}[1]
    \vspace{0.0in}
    \STATE $\bx_T = \mathbf{y}_{\sf r}$ 
    \FOR{$t=T, ... , 1$}
      \STATE $\bz \sim \mathcal{N}(\bzero_{2K}, \bI_{2K})$ if $t > 1$, else $\bz = \bzero_{2K}$
      \STATE $\bx_{t-1} = \frac{1}{\sqrt{\alpha_t}}\left( \bx_t - \frac{1-\alpha_t}{\sqrt{1-\bar\alpha_t}} \bepsilon_{\bm \theta}(\bx_t, t) \right) + \sqrt{1-\alpha_t} \bz$
    \ENDFOR
    \STATE \textbf{return} $\bx_0$
  \end{algorithmic}
\end{algorithm}
\vspace{-8mm}
\end{figure}

\vspace{0mm}
\section{Evaluations}\label{sec:Eval}
\begin{figure}
\centering 
    \begin{subfigure}{0.48\textwidth}
        \centering
        \includegraphics[width=\linewidth, trim={12.0in 0.0in 2.7in  0.15in},clip]{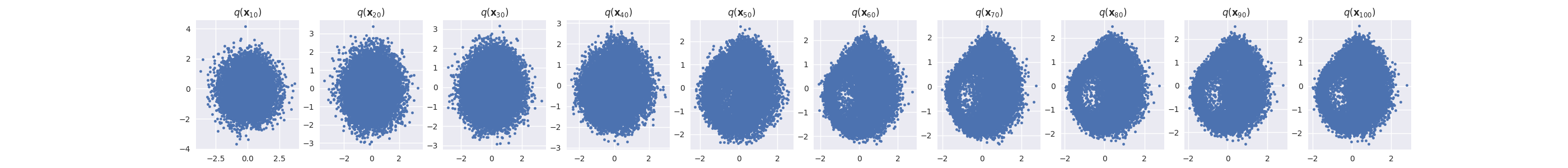}
    \end{subfigure}
    \begin{subfigure}{0.48\textwidth}
        \centering
        \includegraphics[width=\linewidth, trim={12.0in 0.0in 2.7in  0.0in},clip]{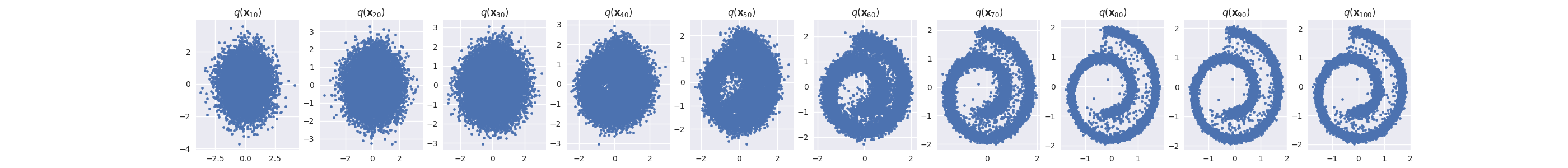}
    \end{subfigure}
    \begin{subfigure}{0.48\textwidth}
        \centering    
        \includegraphics[width=\linewidth, trim={12.0in 0.0in 2.7in  0.0in},clip]{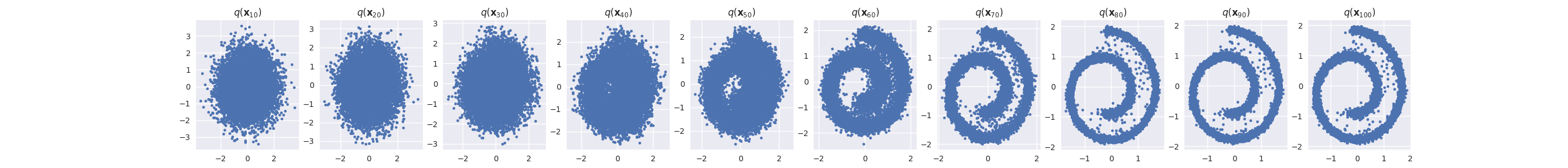}
    \end{subfigure}
    \caption{\small Data visualization for the training process. The rows correspond to epochs $400$, $1200$, and $2000$, respectively. }
    \label{fig:train}
    \vspace{-2mm}
\end{figure}

\begin{figure} 
\centering
\includegraphics
[width=3.5in,height=2.35in,trim={0.2in 0.2in 0.3in  0.5in},clip]{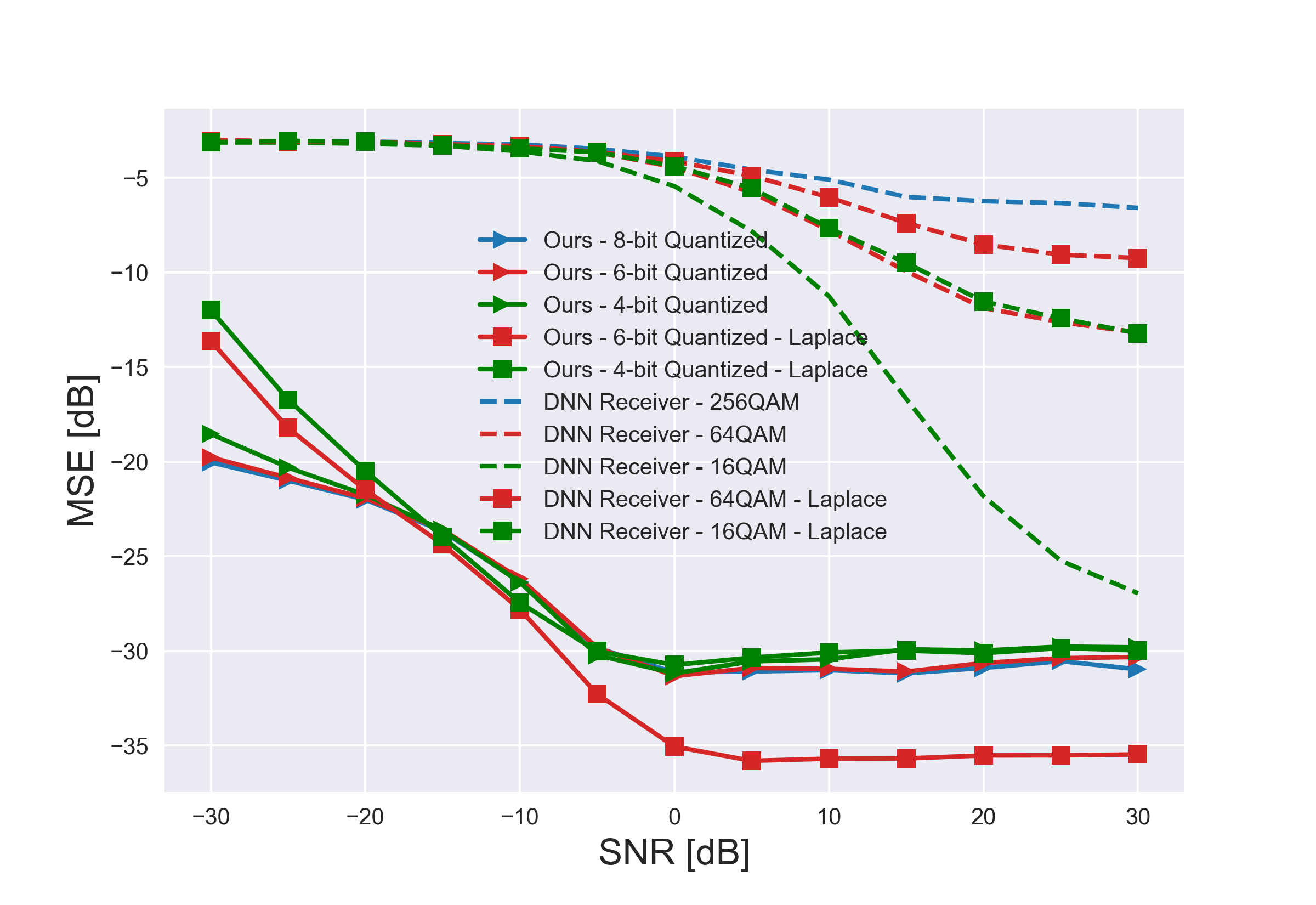}
\vspace{-3mm}
\caption{\small MSE between the original signal and the reconstructed one  under AWGN channel and non-Gaussian additive noise.  
}
	\label{fig:mse}
 \vspace{-5mm}
\end{figure}

\begin{figure} 
\centering
\includegraphics
[width=3.4in,height=2.2in,trim={0.0in 0.0in 0.0in  0.0in},clip]{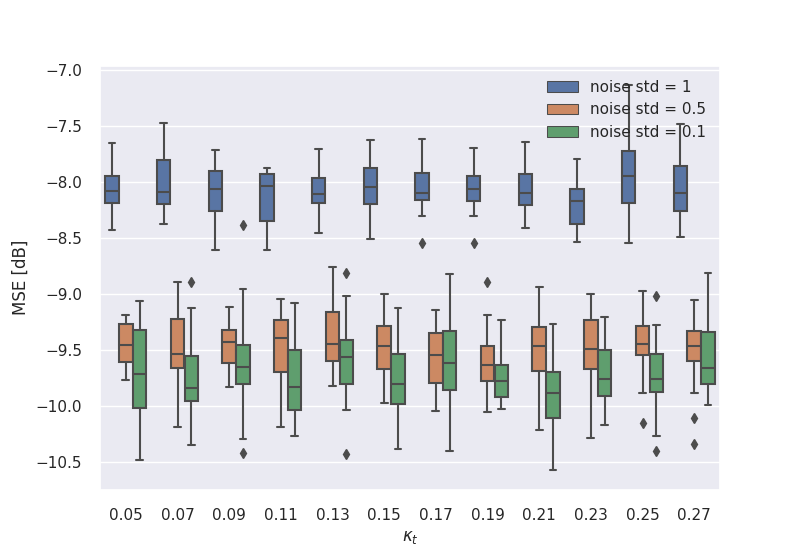}
\vspace{0mm}
\caption{\small MSE between the original signal and the reconstructed one  for different levels of hardware impairment over fading channels.} 
	\label{fig:boxplot}
 \vspace{-5mm}
\end{figure}

\vspace{0mm}
In this section, we provide  numerical results  to highlight the performance of the proposed scheme under AWGN and fading channels.  We  show that the DDPM method  can provide  \emph{resilience} for the communication system under low-SNR regimes, non-Gaussian additive noise, and  quantization errors.   
We also  compare the  performance of our DDPM-based scheme to the  DNN-based receiver  of \cite{DNN} as one of the seminal  benchmarks for ML-based  communications systems.      
For training the diffusion model, we use adaptive moment estimation (Adam) optimizer with  learning rate  $\lambda = 10^{-3}$ over $2000$ epochs, i.e., the stopping criterion in Algorithm \ref{alg:trainAlg} is reaching the maximum number of epochs \cite{CDDM, DM_for_E2EComm, CGM_ChanEst}.   
Transmit SNR is defined as 
 $   \Gamma = 10 \log_{10} (\frac{p}{\delta^2}) \hspace{1mm}\text{dB}.$ 
Without loss of generality, we set the average signal power to $p = 1$ for all experiments, and  vary the
SNR by setting the standard deviation (std)  of noise $\delta$ \cite{AE_paper}.   
For Figs.  \ref{fig:train} and \ref{fig:mse}, we set $\kappa^t = 0.05$ and $\kappa^r = 0.15$ \cite{HI}. We  study the effect of different HWI levels in Fig. \ref{fig:boxplot}.   

\vspace{0mm}
Fig. \ref{fig:train} visualizes the denoising and  generative performance of the implemented  diffusion model during training over swiss roll dataset with $10000$ samples.     
For this figure, we took ``snapshots''   by  saving the model's current state  at specific checkpoints (every $400$ epochs) during the training process,    
and the corresponding  output of the DDPM  is plotted  over time-steps $t = \{50, 60, \ldots, 100\}$. 
As can be seen from the figure, our DDPM gradually learns to denoise and generate samples out of an isotropic Gaussian distribution, where  data samples with different levels of ``noisiness'' can be observed  at  different time-steps $t \in [T]$ from left to right. Moreover, as we reach the maximum number of  epochs, the model can sooner (i.e., in fewer time-steps) generate samples.

We now study the   performance of our system under low-SNR regimes, non-Gaussian additive noise, and  quantization errors.  
Fig. \ref{fig:mse} demonstrates the reconstruction performance of our DDPM-based  scheme compared to the conventional DNN-based benchmark \cite{DNN}  over a wide range of SNR values.  
For this experiment,  the original data samples are first qunatized into bitstreams, and then mapped  to QAM  symbols   (as a   widely-adopted constellation format in wireless networks \cite{twelve_6G, hexaX}) for transmission.  
The MSE metric (averaged over $10$ runs of sampling) between the original and  the reconstructed data samples is considered  for evaluation.

Since we have employed  our DDPM at the receiver,  
we only exploit the receiver DNN of \cite{DNN}  and fine-tune it for benchmarking, in order to  have a fair comparison.    
For the DNN benchmark, three linear layers with $64$ neurons at hidden layers and rectified linear unit (ReLU) activation functions are considered.  
For $16$- and $64$-QAM scenarios, we considered $5000$ training iterations, while for $256$-QAM, the DNN was trained for $30000$ iterations with Adam optimizer and learning rate $\lambda = 0.01$.   
Notably, Fig. \ref{fig:mse}   highlights significant improvement in reconstruction performance among different quantization resolutions and  across a wide range of SNR values, specially in low-SNR regimes.  
For instance,  more than $25$ dB improvement is achieved at $0$ dB SNR for $64$-QAM scenario,  compared to the the well-known model of \cite{DNN} as one of the promising benchmarks in ML-based  communications systems.  
Fig. \ref{fig:mse}  also highlights the \emph{resilience} of our approach against non-Gaussian  noise. In this experiment we consider additive Laplacian noise with the same variance as that of AWGN scenario \cite{twelve_6G}.  The non-Gaussian noise can happen due to the non-Gaussian interference in multi user scenarios \cite{twelve_6G}. Remarkably, although we do not re-train our diffusion model under Laplacian noise,  the performance of  our  approach does not change (or even becomes better) under this non-Gaussian assumption. However, we can see from the figure that the DNN benchmark can experience significant  performance degradation under non-Gaussian assumption,  although  we have  re-trained the DNN  with Laplacian noise.     This highlights the \emph{robust out-of-distribution performance} of our proposed scheme.

Fig. \ref{fig:boxplot} studies the effect of different HWI levels on the reconstruction performance of our scheme over Rayleigh fading channels. For this figure, we set $\kappa^t = 0.05$ and vary the impairment level  of the receiver, $\kappa^r$,  over the typical ranges specified in  \cite{HI}. The reconstruction results are obtained in terms of MSE metric over $20$ realizations of the system.  The figure  highlights an important characteristic  of our proposed scheme. Our DDPM-based communication system  is \emph{resilient} against hardware and channel distortions, as the reconstruction performance does not change with the increase in the impairment  level.  Notably, our DDPM-based scheme showcases a \emph{near-invariant} reconstruction performance with respect to channel  noise and  impairment levels, which also highlights the generalizability of the proposed approach.    This  is achieved  due to the carefully-designed  so-called ``variance scheduling''  of our DDPM framework in \eqref{eq:fwd_sample_gen_diffusion}, which  allows the system to become robust against a wide range of distortions caused by channel and hardware impairments.

 \vspace{0mm}
\section{Conclusions}\label{sec:concl} 
\vspace{-1mm}
In this paper,  we have studied  the application of DDPMs  in wireless communication  systems under   HWIs. After introducing  the DDPM framework and formulating our system model, we have evaluated the reconstruction  performance of our scheme in terms of reconstruction performance. 
We have demonstrated the  \emph{resilience} of our  DDPM-based approach under   low-SNR regimes, non-Gaussian noise, and   different HWI levels. Our results have shown that more than $25$ dB improvement in MSE is achieved compared to DNN-based receivers, 
as well as robust out-of-distribution performance.

\def\baselinestretch{0.98}


\end{document}